\documentclass[12pt]{article}

\usepackage{graphicx}

\begin{document}

\begin{center}

{\Large \bf Chiral transition and deconfinement transition in \\[2mm]
QCD with the highly improved staggered quark\\[2mm]
(HISQ) action}\\[3mm]

Alexei Bazavov$^a$\footnote{Present address: Physics Department, Brookhaven National Laboratory, 
         Upton, NY 11973, USA}  and Peter Petreczky$^b$ 

        (for HotQCD collaboration)\footnote{
The HotQCD Collaboration members are:
A.~Bazavov,
T.~Bhattacharya,
M.~Cheng,
N.H.~Christ,
C.~DeTar,
S.~Gottlieb,
R.~Gupta,
U.M.~Heller,
C.~Jung,
F.~Karsch,
E.~Laermann,
L.~Levkova,
C.~Miao,
R.D.~Mawhinney,
S.~Mukherjee,
P.~Petreczky,
D.~Renfrew,
C.~Schmidt,
R.A.~Soltz,
W.~Soeldner,
R.~Sugar,
D.~Toussaint,
W.~Unger
and
P.~Vranas
} \\[2mm]

$^a$Department of Physics, University of Arizona,
         Tucson, AZ 85721, USA

$^b$Physics Department, Brookhaven National Laboratory, 
         Upton, NY 11973, USA

\end{center}

\begin{abstract}
We report preliminary results on the chiral and deconfinement aspects of the QCD transition at finite
temperature using the Highly Improved Staggered Quark (HISQ) action on lattices 
with temporal extent of $N_{\tau}=6$ and $8$. The chiral aspects of the transition are studied
in terms of quark condensates and the disconnected chiral susceptibility.
We study the deconfinement transition in terms of the strange quark number susceptibility
and the renormalized Polyakov loop. We made continuum estimates 
for some quantities and find reasonably good agreement between our results
and the recent continuum extrapolated results obtained with the stout staggered quark action.
\end{abstract}

\section*{Introduction}

Improved staggered fermion formulations are widely used to study 
QCD at non-zero temperatures and densities, see e.g. Ref. \cite{carleton,petr}
for recent reviews, for, at least, two reasons:
they preserve a part of the chiral symmetry
of continuum QCD which allows one
to study the chiral aspects of the finite temperature transition,
and are relatively inexpensive to simulate numerically 
because due to the absence of an additive mass
renormalization the Dirac operator is bounded from below.
However, lattice artifacts related to taste symmetry breaking turned
out to be numerically large.
To reduce the taste violations smeared links,
i.e. weighted averages of different paths on the lattice that connect
neighboring points, are used in the staggered
Dirac operator and several improved staggered formulations, like
p4, asqtad, stout and HISQ differ in the choice
of the smeared gauge links. The ones in the p4 and asqtad actions
are linear combinations of single links and different staples 
\cite{karsch01,orginos} and therefore are not elements of the SU(3) group.

It is known that projecting the smeared gauge fields onto the
SU(3) group greatly improves the taste symmetry \cite{anna}. The
stout action \cite{fodor05} and the HISQ action implement the
projection of the smeared gauge field onto the SU(3) (or simply U(3)) group 
and thus achieve better
taste symmetry at a given lattice spacing. For studying QCD at high
temperature it is important to use discretization schemes which 
improve the quark dispersion relation, thus eliminating the 
tree level ${\cal O}(a^2)$ lattice artifacts in thermodynamic quantities.
The p4 and asqtad actions implement this improvement by introducing
3-link terms in the staggered Dirac operator.

In this paper we
report preliminary results on the chiral and deconfinement transition in QCD at non-zero
temperature obtained with the
HISQ action which combines the removal of tree level ${\cal O}(a^2)$ lattice artifacts
with the addition of projected smeared links that greatly improve the taste symmetry.
We also compare our results to the continuum extrapolated results obtained with
the stout action \cite{fodor10}.

\section{Action and run parameters}

The Highly Improved Staggered Quark (HISQ) action developed 
by the HPQCD/UKQCD collaboration \cite{Follana:2006rc}
reduces taste symmetry breaking and decreases the splitting
between different pion tastes by a factor of about three 
compared to the
asqtad action. The net result, as recent scaling studies show
\cite{Bazavov:2009wm,MILC_hisq_2009}, 
is that a HISQ ensemble at lattice
spacing $a$ has scaling violations comparable to ones 
in an asqtad ensemble at lattice spacing $2a/3$.

In this study we used the HISQ action 
in the fermion sector
and the tree-level Symanzik improved gauge action. 
The strange quark mass $m_s$ 
was set to its physical value adjusting 
the quantity $\sqrt{2m_K^2-m_\pi^2}=m_{\eta_{s\bar s}}\simeq \sqrt{2B m_s}$ 
to the physical  value 686.57~MeV. We used two values of the light quark mass:
$m_l=0.05m_s$ and $m_l=0.20m_s$. 
These correspond to the lightest pion mass of about $160$~MeV and $320$~MeV
respectively \cite{wwnd10}.      
Calculations
have been performed on $24^3 \times 6$ and $32^3 \times 8$ lattices
for the smaller quark mass, while for the larger quark mass we used $16^3 \times 6$ lattices.
At several values of the lattice spacings zero temperature simulations
have been performed on $32^4$ lattices for $m_l=0.05m_s$ and on $16^3 \times 32$ lattices for $m_l=0.2m_s$.
The molecular dynamics (MD)
trajectories have length of 1 time unit (TU) and
the measurements were performed every 5 TUs at zero
and 10 TUs at finite temperature. 
However, for the few smallest beta values the trajectories had length of $1/2$ or $1/3$ TU. In this
case measurements have been performed after each 20 or 30 trajectories (which corresponds, again, to 10 TUs).
Typically, at least
300 TUs were discarded for equilibration at the beginning of the simulations.

The lattice
spacing has been determined by measuring the static
quark anti-quark potential.
As in previous studies by the MILC collaboration the static potential was calculated fixing to Coulomb gauge
and considering temporal Wilson lines of different extent. Forming the ratio of these correlators
and fitting them to a constant plus linear Ansatz we extracted the static potential $V(r)$. We have calculated
the Sommer scale $r_0$ defined as
\begin{equation}
\left.r^2 \frac{d V}{dr}\right|_{r=r_0}=1.65.~~~~~
\end{equation}
The potential has been normalized to the string potential

\begin{equation}
V_{string}=-\frac{\pi}{12r}+\sigma r,
\end{equation}
at $r=1.5r_0$ or equivalently to the value $0.91/r_0$ at $r=r_0$.
The additive constant determined by this normalization is used to
calculate the renormalization constant for the Polyakov loop as will be discussed later.
To convert from lattice units to physical units we use the value 
$r_0=0.469$ fm as determined in \cite{gray}.

\section{Chiral and deconfinement transition in QCD}

\subsection{The chiral transition}

\begin{figure}
\includegraphics[width=0.50\textwidth]{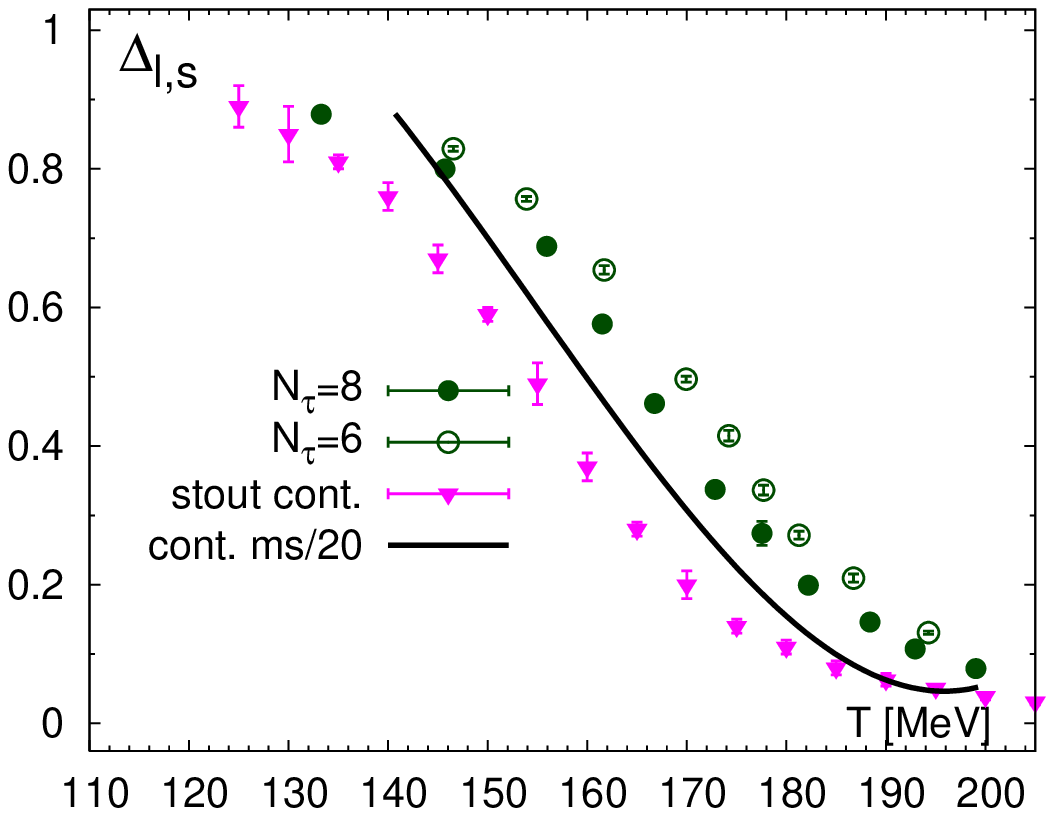}
\includegraphics[width=0.50\textwidth]{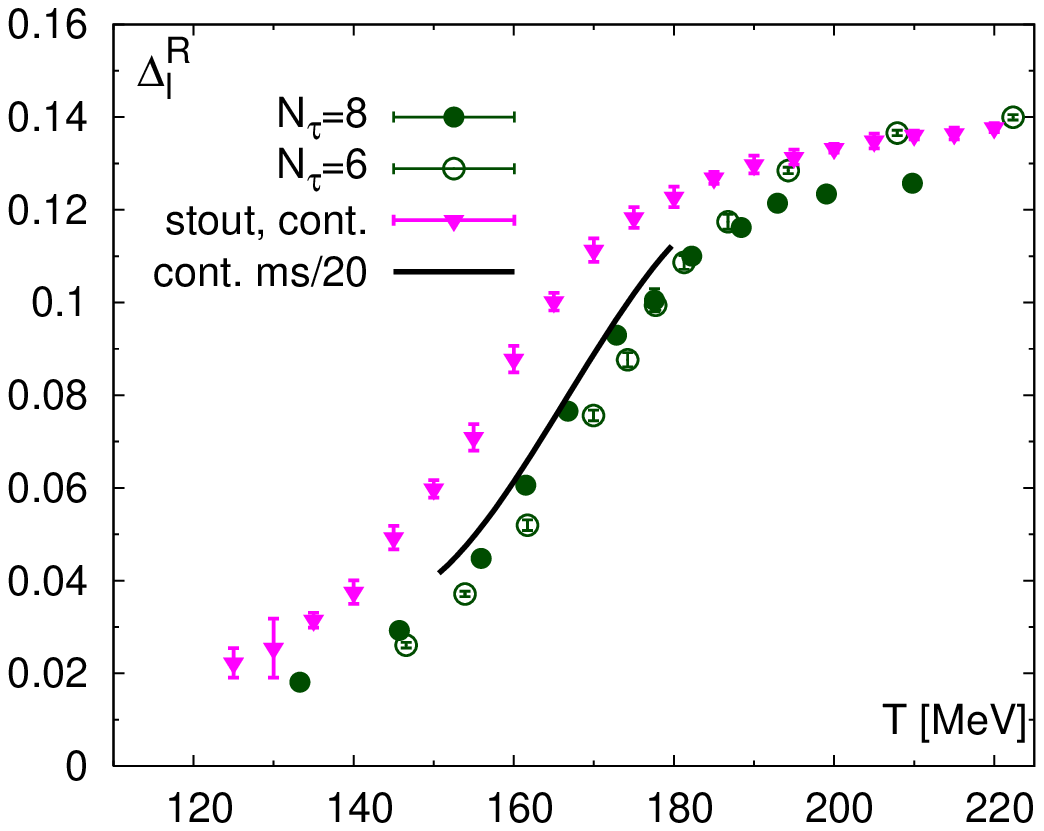}
\caption{The subtracted chiral condensate and $\Delta_l^R$ compared to 
the continuum estimate for the stout action \cite{fodor10}. The solid
black line is our continuum estimate for the HISQ action.
}
\label{fig:pbp}
\end{figure}

In the limit of zero light quark masses QCD has a chiral symmetry and
the finite temperature transition is a true phase transition. The order
parameter for this transition is the light chiral condensate 
$\langle \bar \psi  \psi \rangle_{l}$.
However, even at finite values of the quark mass the chiral condensate 
will show a rapid change in the
transition region indicating an effective restoration of chiral symmetry. 
The fluctuations of the chiral condensate, also called the disconnected
chiral susceptibility, will have a peak at the transition temperature.
Recent studies with the p4 action suggest that for the physical strange quark mass
and two light quarks the chiral transition is of the second order for vanishing light quark mass,
belonging to the $O(N)$ universality class \cite{scaling}.
Thus, the universal properties of the chiral transition in the limit
of zero light quark masses govern the transition for sufficiently small
but non-zero light quark masses \cite{scaling}. The corrections to scaling
turned out to be small for $m_l=m_s/20$ and thus the  temperature variation
of the chiral condensate is to a large extent determined by the singular part of the free
energy density which is universal \cite{scaling}. This allows to determine the
chiral transition temperature for non-zero quark masses. The temperature
derivative of the chiral condensate and disconnected chiral susceptibility 
will diverge in the limit $m_l\rightarrow 0$.
Since the chiral condensate
has an additive ultraviolet renormalization we consider 
the subtracted chiral condensate \cite{rbcbi}
\begin{equation}
\displaystyle
\Delta_{l,s}(T)=\frac{\langle \bar\psi \psi \rangle_{l,\tau}-\frac{m_l}{m_s} \langle \bar \psi \psi \rangle_{s,\tau}}
{\langle \bar \psi \psi \rangle_{l,0}-\frac{m_l}{m_s} \langle \bar \psi \psi \rangle_{s,0}}.
\end{equation}
Here the subscript $l$ and $s$ refer to light and strange chiral condensates, while subscript
$0$ and $\tau$ to the expectation value at zero and non-vanishing temperature.
Another possibility to get rid of the additive renormalization in the chiral condensate 
is to consider the quantity
\begin{equation}
\Delta_l^R(T)=-m_s r_0^4 (\langle \bar\psi \psi \rangle_{l,\tau}-\langle \bar \psi \psi \rangle_{l,0})
\end{equation}
which up to a constant multiplicative factor is identical to the quantity introduced in Ref. \cite{fodor10} and
was called the renormalized chiral condensate.
In Fig. \ref{fig:pbp} we show our numerical results for $\Delta_{l,s}$ and $\Delta_l^R$ and compare
them to the continuum extrapolated stout results \cite{fodor10}. To facilitate this comparison
for $m_l=0.05m_s$ we performed a combined polynomial fit of the $N_{\tau}=6$ and $N_{\tau}=8$ results 
allowing for lattice spacing dependent coefficients in this fit. This allows us to give an estimate of these
quantities in the continuum limit shown as the solid black lines. The continuum estimates for HISQ
are different from the continuum extrapolated stout results. This is presumably due to the fact that
the quark masses used in the stout calculations are smaller, namely $m_l=m_s/27.3$ \cite{fodor09}.
The transition temperature is decreasing with decreasing the quark mass. Also the chiral condensate
at fixed temperature decreases with decreasing the quark mass. Thus at a qualitative level this discrepancy
could be understood. Future calculations on $N_{\tau}=12$ lattices and smaller quark masses will
be needed to clarify this issue quantitatively.       
\begin{figure}
\includegraphics[width=0.49\textwidth]{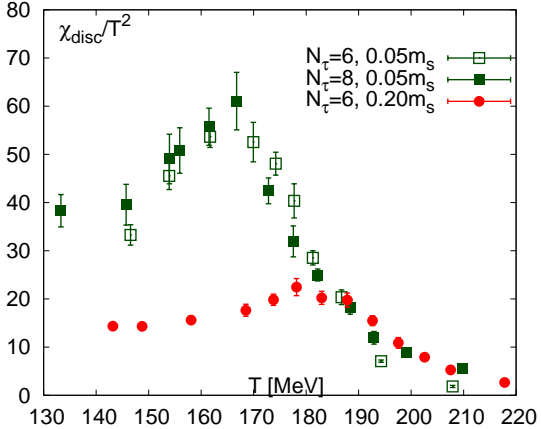}\hfill
\caption{
The disconnected chiral susceptibility calculated with the HISQ action on $N_{\tau}=6$
and $8$ lattices for $m_l=0.05m_s$ and $m_l=0.20m_s$.
}
\label{fig:chi_hisq}
\end{figure}

We also calculated the disconnected chiral susceptibility.
In Fig. \ref{fig:chi_hisq} we show the  light  quark disconnected chiral susceptibility. The peak
position in this quantity also defines the chiral transition temperature. To estimate the peak
position in this quantity we have performed fits of the lattice data using different functional
forms. We also varied the fit intervals. Our analysis gives
\begin{eqnarray}
&&
N_{\tau}=6:~~~T_c=168(4)(5){\rm MeV}~~(0.05m_s);~~~~~~~ T_c=185(4)(5){\rm MeV}~~(0.20m_s)\\
&&
N_{\tau}=8:~~~T_c=165(4)(5){\rm MeV}~~(0.05m_s),
\end{eqnarray}
where the first error reflects the uncertainty in the determination of the peak
position estimated using different fit forms and varying the fit range. The second error
is due to the scale determination. The peak positions in the disconnected chiral susceptibilities 
are significantly lower than those obtained with p4 and asqtad actions for $N_{\tau} \le 8$ \cite{rbcbi06,dpf09}, though
they are compatible with the recent estimates for asqtad on $N_{\tau}=12$ lattices \cite{wwnd10}.
Unfortunately no published stout data are available for the disconnected chiral susceptibility
and direct comparison between HISQ and stout results is not possible here.
Therefore, in Ref. \cite{wwnd10} we compared HISQ and stout data for the renormalized chiral
susceptibility introduced in \cite{fodor06}. Some small discrepancies between the stout and HISQ
results have been found there which again are probably due to slightly different quark masses.
Note that within errors of the calculations the peak positions in the renormalized chiral susceptibility
are consistent between HISQ and stout \cite{wwnd10}.

\subsection{Deconfinement transition}
\begin{figure}
\includegraphics[width=0.5\textwidth]{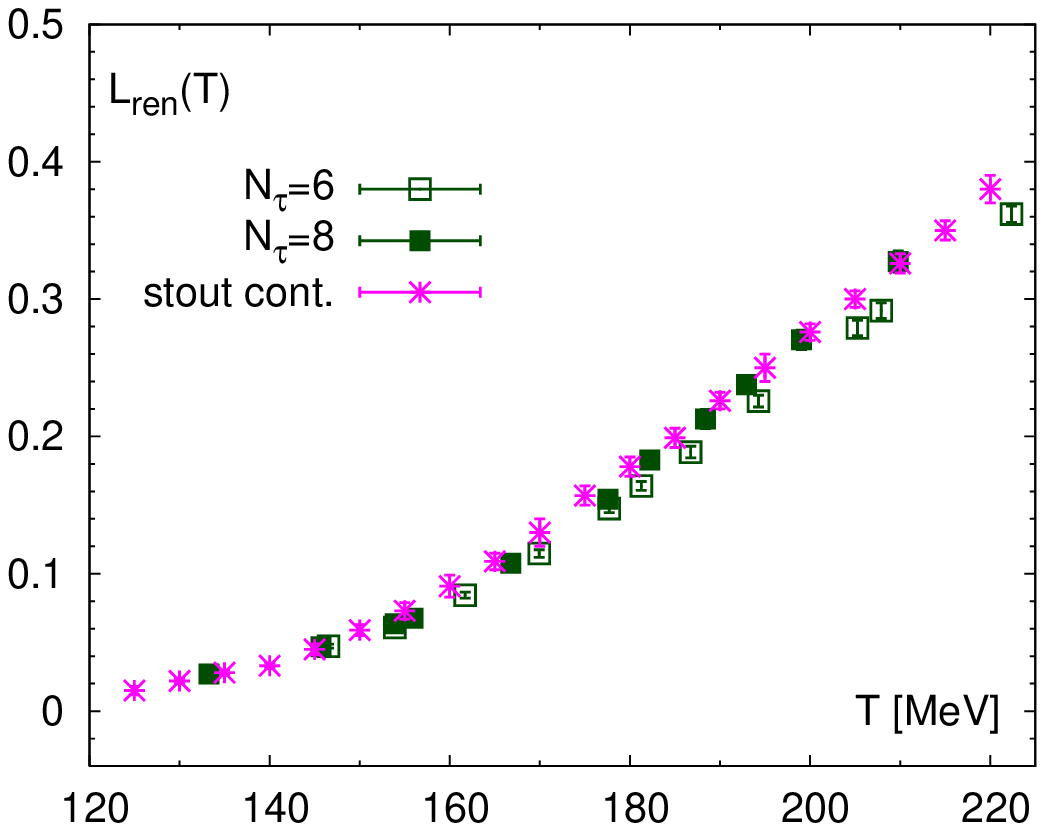}
\includegraphics[width=0.5\textwidth]{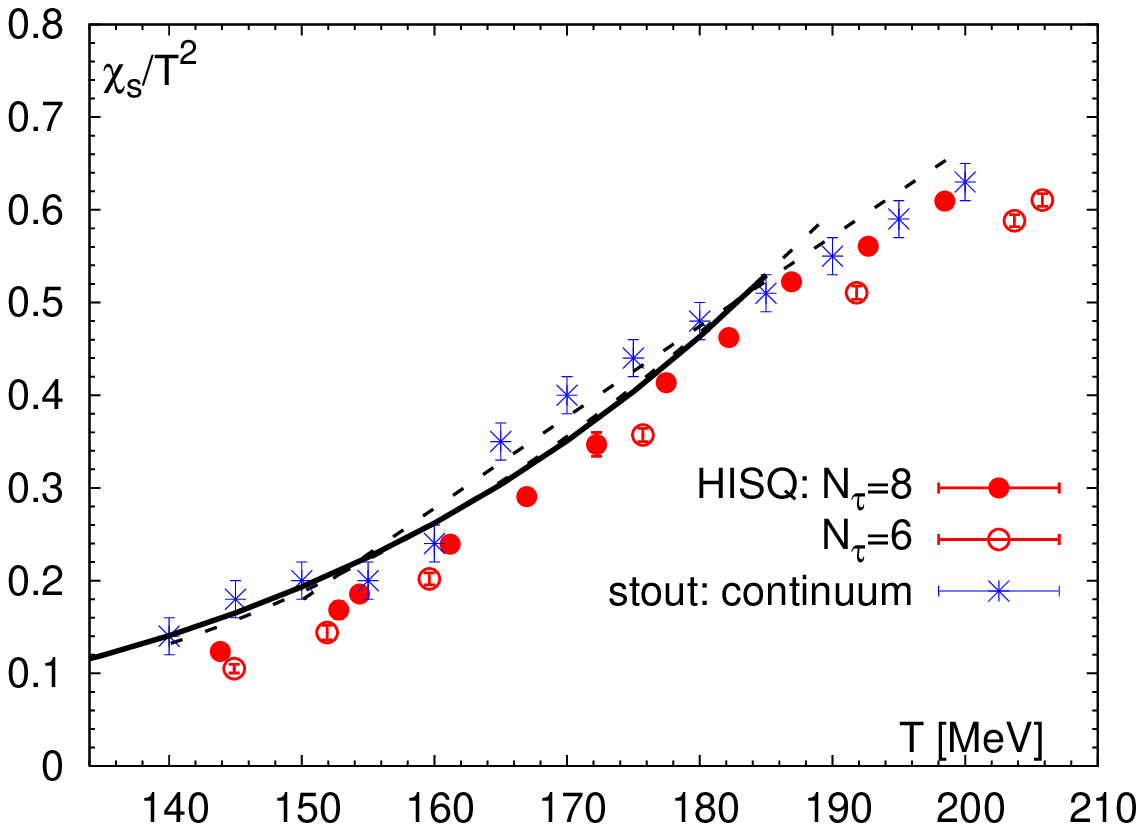}
\caption{The renormalized Polyakov loop (left) and the strange quark number susceptibility
(right) calculated with the HISQ action for $m_l=0.05m_s$ and compared to the continuum extrapolated stout results.
The lines correspond to continuum extrapolation.
}
\label{fig:poly}
\end{figure}

By the deconfinement transition we mean liberation of many degrees of freedom, which
also could be understood as a transition from hadronic degrees of freedom to partonic
ones, and the onset of color screening.
The aspects of the deconfinement transition related to color screening are studied
in terms of the Polyakov loop and Polyakov loop correlators.
The Polyakov loop needs to be renormalized and after proper renormalization it is
related to the free energy of a static quark anti-quark pair at infinite separation~$F_{\infty}(T)$
\cite{okacz02},
\begin{equation}
L_{ren}(T)=\exp(-F_{\infty}(T)/(2 T)).
\end{equation}
The renormalized Polyakov loop can be obtained from the bare Polyakov loop as
\begin{eqnarray}
&
\displaystyle
L_{ren}(T)=z(\beta)^{N_{\tau}} L_{bare}(\beta)=
z(\beta)^{N_{\tau}} \left<\frac{1}{3}  {\rm Tr } 
\prod_{x_0=0}^{N_{\tau}-1} U_0(x_0,\vec{x})\right >.
\end{eqnarray}
Here the multiplicative renormalization constant $z(\beta)$ is related to the 
additive normalization of the potential $c(\beta)$ as $z(\beta)=\exp(-c(\beta)/2)$
discussed in the previous section.
The renormalized Polyakov loop has been calculated for pure gauge theory \cite{okacz02,digal03},
3-flavor QCD \cite{petrov04} as well as for 2-flavor QCD \cite{okacz05}. More recently it has been
calculated for 2+1 flavor QCD with physical strange quark mass and light quark masses close
to the physical values \cite{fodor10,rbcbi,hotqcd,eos005,fodor06,fodor09}. 
In Fig. \ref{fig:poly} we show our results for the renormalized Polyakov loop calculated
with the HISQ action using $N_{\tau}=6$ and $8$ lattices and $m_l=0.05m_s$ and compare them to the continuum
extrapolated results obtained with the stout action \cite{fodor10}.
The decrease of $F_{\infty}(T)$, and thus the increase in the Polyakov loop could be related 
to the onset of screening at high temperatures
(e.g. see discussion in Ref. \cite{petr_hp04}). On the other hand, in the 
low-temperature region the increase of $L_{ren}$ is related to the fact that there are many static-light meson
states that can contribute to the static quark free energy close to the transition temperature, while
far away from the transition temperature it is determined by the binding energy of the lowest static-light mesons.
The strange quark number susceptibility is defined as the second derivative of the pressure with respect
to the quark chemical potential
\begin{equation}
\chi_s=\frac{\partial^2 p(T,\mu_s)}{\partial \mu_s^2} |_{\mu_s=0}.
\end{equation}

It describes strangeness fluctuations at zero strange quark chemical potential.
At low temperatures strangeness is carried by strange hadrons, which are heavy compared
to the temperature. As a result strangeness fluctuations are suppressed in the low-temperature
region. At high temperatures, on the other hand, strangeness is carried by light
quarks and strangeness fluctuations are close to the value given by an ideal quark gas.
Deconfinement will manifest itself as a rapid increase in the strangeness fluctuations in
some temperature interval, reflecting the change in the relevant degrees of freedom from
hadronic to partonic. Therefore, strangeness fluctuations are used as a probe of the deconfinement
transition \cite{fodor10,rbcbi,hotqcd,eos005,fodor06,fodor09}. In Fig. \ref{fig:poly} we show
the strange quark number susceptibility calculated for the HISQ action on $N_{\tau}=6$ and $8$ lattices
for $m_l=0.05m_s$.

We performed a combined polynomial fit of our lattice results where the coefficient of the polynomial had
$a^2$ correction. This allows to estimate strangeness fluctuations in the  continuum limit. We compared
our results to those obtained with the stout action and extrapolated to the continuum. 
As one can see there is reasonable agreement between HISQ and stout results in the continuum limit.
The main difference is the inflection point in the stout calculations at temperatures of about $165$ MeV
which is not visible in the HISQ results. Note, that strangeness fluctuations approach the continuum
limit from below. This is expected due to the lattice spacing dependence of the hadron masses \cite{pasi}.

\section{Conclusions}
We studied the deconfinement and chiral transition in QCD at non-zero temperature using the HISQ
action and compared our results to recent continuum extrapolated results obtained with the stout action.
We find reasonable agreement in the continuum limit between the results obtained
with different actions for the renormalized Polyakov loop and the strange quark number susceptibility.
We also calculated the chiral condensate and disconnected chiral susceptibility for the HISQ action.
For these quantities a direct comparison with stout results is more difficult due the slightly different
light quark masses used in the two calculations. Nonetheless, current HISQ results indicate a chiral
transition temperature that is significantly lower than the previous estimates obtained with the asqtad
and p4 actions using lattices with the temporal extent $N_{\tau} \le 8$.
It would also be interesting to compute the equation of state with the HISQ
action and compare it to the very recent results obtained with the stout action \cite{fodor10eos}.

\section*{Acknowledgements}
 This work has been supported in part by contracts DE-AC02-98CH10886
 and DE-FC02-06ER-41439 with the U.S. Department of Energy
 and contract 0555397 with the National Science Foundation. The numerical calculations have been performed
 using the USQCD resources at Fermilab as well as the BlueGene/L
 at the New York Center for Computational Sciences (NYCCS).

\end{document}